\documentclass[conference,comsoc]{IEEEtran}
\usepackage[pdftex]{graphicx}
\usepackage{epstopdf}
\usepackage[T1]{fontenc}
\usepackage[cmex10]{amsmath}
\usepackage{ifthen}
\usepackage{commath,bm}
\usepackage[cmintegrals]{newtxmath}
\usepackage{esint}
\usepackage{cite} 
\usepackage{hyperref}
\usepackage{multirow}
\usepackage{xcolor}
\newtheorem{theorem}{Theorem}
\newtheorem{corollary}{Corollary}
\newtheorem{remark}{Remark}

\newcommand{\pr}{\text{Pr}}

\newcommand{\Pe}{\mathrm{P_e}}

\newcommand{\seq}{\mathbf{b}}

\newcommand{\bit}{b}
\newcommand{\diffd}[1]{D_{#1}} 
\newcommand{\TX}{\text{tx}}
\newcommand{\RX}{\text{rx}}
\newcommand{\hs}{\hspace{-0.8 mm}}

\allowdisplaybreaks
\newcounter{storeeqcounter}
\newcounter{tempeqcounter}
\begin{document}

\title{Statistical Analysis of Time-Variant Channels in Diffusive Mobile Molecular Communications\vspace{-3 mm}}
\author{
\IEEEauthorblockN{Arman~Ahmadzadeh, Vahid~Jamali, and~Robert~Schober}
\IEEEauthorblockA{Friedrich-Alexander-University Erlangen-Nuremberg , Germany \vspace{-4 mm}}
}

\maketitle 
\begin{abstract} 
In this paper, we consider a diffusive mobile molecular communication (MC) system consisting of  a pair of mobile transmitter and receiver nano-machines suspended in a fluid medium, where we model the mobility of the nano-machines by Brownian motion. The transmitter and receiver nano-machines exchange information via diffusive signaling molecules. Due to the random movements of the transmitter and receiver nano-machines, the statistics of the channel impulse response (CIR) change over time. We introduce a statistical framework for characterization of the impulse response of time-variant MC channels. In particular, we derive closed-form analytical expressions for the mean and the autocorrelation function of the impulse response of the channel. Given the autocorrelation function, we define the coherence time of the time-variant MC channel as a metric that characterizes the variations of the impulse response. Furthermore, we derive an analytical expression for evaluation of the expected error probability of a simple detector for the considered system. In order to investigate the impact of CIR decorrelation over time, we compare the performances of a detector with perfect channel state information (CSI) knowledge and a detector with outdated CSI knowledge. The accuracy of the proposed analytical expression is verified via particle-based simulation of the Brownian motion.          
\end{abstract}

\section{Introduction}
Future synthetic nano-networks are expected to facilitate new revolutionary applications in areas such as biological engineering, healthcare, and environmental engineering \cite{NakanoB}. Molecular communication (MC), where molecules are the carriers of information, is one of the most promising candidates for enabling reliable communication between nano-machines in such future nano-networks due to its bio-compatibility, energy efficiency, and abundant use in natural biological systems. 

Some of the envisioned application areas for synthetic MC systems may require the deployment of \emph{mobile} nano-machines. For instance, in targeted drug delivery and intracellular therapy applications, it is envisioned that mobile nano-machines carry drug molecules and release them at the site of application, see \cite[Chapter 1]{NakanoB}. As another example, in molecular imaging, a group of mobile bio-nano-machines such as viruses carry green flurescent proteins (GFPs) to gather information about the environmental conditions from a large area inside a targeted body, see \cite[Chapter 1]{NakanoB}. In order to establish a reliable communication link between nano-machines, knowledge of the channel statistics is necessary. However, for \emph{mobile} nano-machines these statistics change with time, which makes communication even more challenging. Thus, it is crucial to develop a mathematical framework for characterization of the stochastic behaviour of the channel. Stochastic channel models provide the basis for the design of new modulation, detection, and/or estimation schemes for mobile MC systems. 

In the MC literature, the problem of mobile MC has been considered in \cite{108-Luo2016, 109-Qiu2016, 106-Hsu2015, 116-Jamali2016, 105-Guney2012, 107-Kuscu2014, 104-Nakano2016}. However, none of the previous works provided a stochastic framework for the modeling of time-variant channels. In particular, in \cite{108-Luo2016, 106-Hsu2015, 109-Qiu2016, 116-Jamali2016} it is assumed that \emph{only} the receiver node is mobile and the channel impulse response (CIR) either changes slowly over time, due to the slow movement of the receiver, as in \cite{109-Qiu2016}, or it is fixed for a block of symbol intervals and may change slowly from one block to the next; see \cite{106-Hsu2015, 116-Jamali2016}. In \cite{105-Guney2012} and \cite{107-Kuscu2014}, a \emph{three-dimensional} random walk model is adopted for modeling the mobility of nano-machines, where it is assumed that information is \emph{only} exchanged upon the collision of two nano-machines. In particular, \emph{F\"orster resonance energy transfer} and a \emph{neurospike communication model} are considered for information exchange between two colliding nano-machines in \cite{105-Guney2012} and \cite{107-Kuscu2014}, respectively. Recently, the authors of \cite{104-Nakano2016} proposed a leader-follower model for target detection applications in two-dimensional mobile MC systems. Langevin equations are used to describe nano-machine mobility and a \emph{non-diffusion} approach is adopted for communication between leader and follower nano-machines. In our previous work \cite{ArmanJ3}, unlike \cite{108-Luo2016, 109-Qiu2016, 106-Hsu2015, 116-Jamali2016, 105-Guney2012, 107-Kuscu2014, 104-Nakano2016}, we have established the mathematical basis required for analyzing mobile MC systems. We have shown that by appropriately modifying the diffusion coefficient of the signaling molecules, the CIR of a mobile MC system can be obtained from the CIR of the same system with fixed transmitter and receiver.

In this paper, we consider a three-dimensional diffusion model to characterize the movements of both transmitter and receiver nano-machines, where unlike \cite{105-Guney2012, 107-Kuscu2014, 104-Nakano2016} we assume that nano-machines exchange information via diffusive signaling molecules. Furthermore, unlike \cite{ArmanJ3}, we develop a \emph{stochastic} framework for describing the time-varying CIR of the mobile MC system. To the best of the authors' knowledge, a stochastic channel model for mobile MC systems has not been reported yet. In particular, this paper makes the following contributions:     

\begin{itemize}
	\item We establish a mathematical framework for the characterization of the time-varying CIR of mobile MC systems as a stochastic process, i.e., we introduce a stochastic channel model.  
	\item We derive closed-form analytical expressions for the first-order (mean) and second-order (autocorrelation function) moments of the time-varying CIR of mobile MC systems.
	\item Equipped with the autocorrelation function of the CIR, we define the coherence time of the channel as the time during which the CIR does not substantially change.
	\item To evaluate the impact of the CIR decorrelation occurring in mobile MC systems on performance, we derive the expected bit error probability of a simple detector for perfect and outdated CSI knowledge, respectively.     
\end{itemize}

The rest of this paper is organized as follows. In Section \ref{Sec.SysMod}, we introduce the system model. In Section \ref{Sec.StoChaMod}, we develop the proposed stochastic channel model, calculate the mean and autocorrelation function of the CIR, and derive the coherence time of the channel. Then, in Section \ref{Sec.PerAna}, we calculate the expected bit error probability of the considered system for detectors with perfect and outdated CIR knowledge, respectively. Simulation and analytical results are presented in Section \ref{Sec.SimRes}, and conclusion are drawn in Section \ref{Sec.Con}.
               
\section{System Model} 
\label{Sec.SysMod} 
\begin{figure}[!t] 
	\centering
	\includegraphics[scale = 0.6]{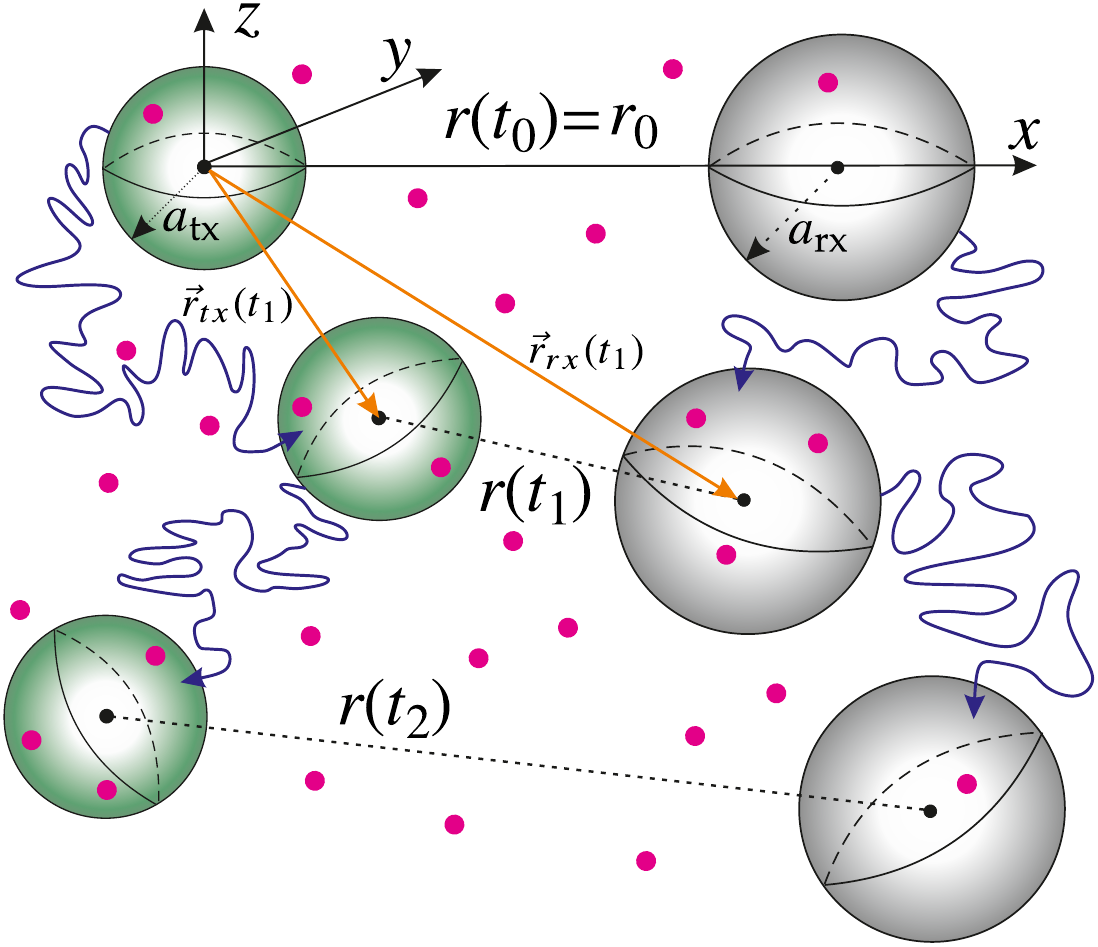}\vspace{-3 mm}
	\caption{Schematic diagram of the considered system model, where the receiver and the transmitter are shown as gray and green spheres, respectively. Sample trajectories of the receiver and the transmitter are shown as blue solid arrows.}\vspace{-5 mm}
	\label{Fig.SystemModel}
\end{figure} 
We consider an unbounded three-dimensional fluid environment with constant temperature and viscosity. The receiver is modeled as a passive observer, i.e., as a transparent sphere with radius $a_{\RX}$ that diffuses with constant diffusion coefficient $\diffd{\RX}$. Furthermore, we model the transmitter as another transparent sphere with radius $a_{\TX}$ that diffuses with constant diffusion coefficient $\diffd{\TX}$. The transmitter employs type $A$ molecules for conveying information to the receiver, which we refer to as $A$ molecules and also as information or signaling molecules. We assume that the $A$ molecules are released in the center of the transmitter and that they can leave the transmitter via free diffusion. In particular, we assume that each signaling molecule diffuses with constant diffusion coefficient $\diffd{A}$ and the diffusion processes of individual $A$ molecules are independent of each other. Moreover, we assume that additional $A$ molecules are uniformly distributed in the environment and impair the reception. These noise molecules may originate from natural sources in the environment.   

Due to the Brownian motion of the transmitter and the receiver, their positions change over time. In particular, we denote the \emph{time-varying} positions of the transmitter and the receiver at time $t$ by $\vec{r}_{\TX}(t)$ and $\vec{r}_{\RX}(t)$, respectively. Then, we define vector $\vec{r}(t) = \vec{r}_{\RX}(t) - \vec{r}_{\TX}(t)$ and denote its magnitude at time $t$ as $r(t)$, i.e., $|\vec{r}(t)| = r(t)$, see Fig.~\ref{Fig.SystemModel}. Furthermore, without loss of generality, we assume that at time $t_0 = 0$ the transmitter and the receiver are located at the origin of the Cartesian coordinate system (i.e., $\vec{r}_{\TX}(t_0=0) = [0,0,0]$) and $\vec{r}_{\RX}(t_0=0) = [x_0,0,0]$, respectively. Thus, $\vec{r}(t_0) = \vec{r}_{\RX}(t_0)$ and $r(t_0 = 0) = r_0 = x_0$.

Furthermore, we assume that the information that is sent from the transmitter to the receiver is encoded into a binary sequence of length $L$, $\seq = [\bit_1, \bit_2, \cdots, \bit_L ]$. Here, $\bit_j$ is the bit transmitted in the $j$th bit interval with $\pr(\bit_j = 1)= P_1$ and $\pr(\bit_j = 0)= P_0 = 1 - P_1$, where $\pr(\cdot)$ denotes probability. We assume that the transmitter and receiver are synchronized, see e.g. \cite{JamaliC2}. We also adopt ON/OFF keying for modulation and a fixed bit interval duration of $T$ seconds. In particular, the transmitter releases a fixed number of $A$ molecules, $N_A$, for transmitting bit ``1'' at the \emph{beginning} of a modulation bit interval and no molecules for transmitting bit ``0''.
\section{Stochastic Channel Model}
\label{Sec.StoChaMod}  
In this section, we first provide some preliminaries regarding the modeling of time-variant channels in diffusive mobile MC systems. Subsequently, we derive a closed-form expression for the autocorrelation function of the impulse response of the considered time-variant channel. Finally, given the derived expression for the autocorrelation function, we define the coherence time of the time-variant MC channel.
\subsection{Impulse Response of Time-Variant MC Channel} 
\begin{figure}[!t] 
	\centering
	\includegraphics[width= 0.49\textwidth, height = 4 cm]{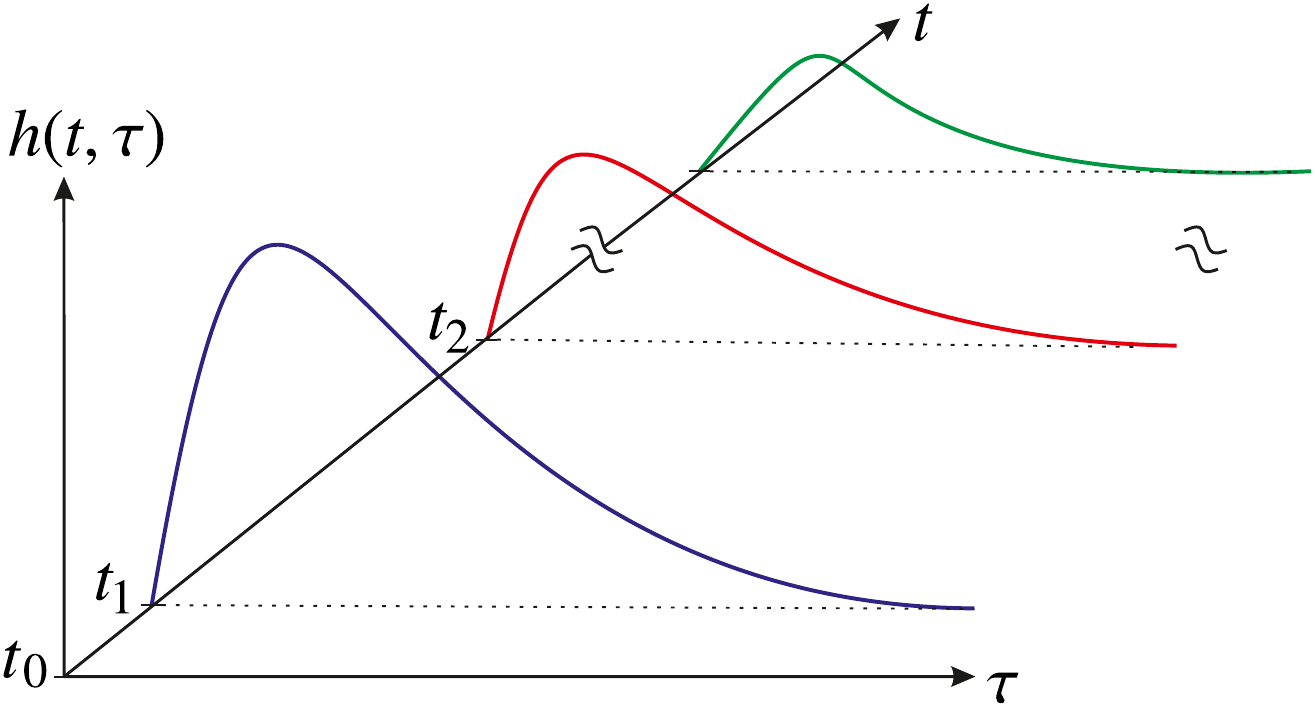}\vspace{-3 mm}
	\caption{Example of CIR variation over time $t$.}\vspace{-5 mm}
	\label{Fig.CIR}
\end{figure}
In this subsection, we first introduce the terminology used for describing the impulse response of the considered time-variant MC channel. Then, subsequently, we present mathematical expressions for the impulse response. We borrow the terminology and the notation for time-variant CIRs from \cite[Ch. 5]{Rappaport}. There, it is assumed that the impulse response of a classical wireless multipath channel can be characterized by a function $h(t,\tau)$, where $t$ represents the time variation due to the mobility of the receiver and $\tau$ describes the channel multipath \emph{delay} for a fixed $t$. Here, we also adopt this notation and derive $h(t,\tau)$ for the problem at hand. In the context of MC, the impulse response of the channel refers to the probability of observing a molecule released by the transmitter at the receiver.  

Let us assume, for the moment, that at the time of release of a given $A$ molecule at the transmitter, $r(t)$ is \emph{known} and given by $r^{\ast}$. Then, the impulse response of the channel, i.e., the probability that a given $A$ molecule, released at the center of the transmitter at time $\tau = 0$, is observed inside the volume of the transparent receiver at time $\tau > 0$ can be written as \cite[Eq. (4)]{NoelJ1}
\begin{IEEEeqnarray}{C}
	\label{Eq.CIRConditioned}
	h(\tau|r^{\ast}) = \frac{V_{\mathrm{obs}}}{(4\pi D_1 \tau)^{3/2}} \exp \left( \frac{- (r^{\ast})^2}{4 D_1 \tau} \right),
\end{IEEEeqnarray}
where $V_{\mathrm{obs}} = \frac{4}{3}\pi a_{\RX}^3$ is the volume of the receiver and $D_1 = \diffd{A} + \diffd{\RX}$ is the effective diffusion coefficient capturing the relative motion of the signaling molecules and the receiver, see \cite[Eq.~(8)]{ArmanJ3}. However, due to the random movements of both the transmitter and the receiver, $\vec{r}(t)$ (and consequently $r(t)$) change randomly. In particular, for the problem at hand, the probability distribution function (PDF) of random variable $\vec{r}(t)$ is given by
\begin{IEEEeqnarray}{C}
	\label{Eq.PDFr(t)} 
	\textit{{\large f}}_{\vec{r}(t)}(\vec{r}) = \frac{1}{(4\pi D_2 t)^{3/2}}\exp \left( \frac{-|\vec{r} - \vec{r}_0|^2}{4 D_2 t} \right),
\end{IEEEeqnarray}
where $D_2 = \diffd{\RX} + \diffd{\TX}$ is the effective diffusion coefficient capturing the relative motion of transmitter and receiver, see \cite[Eq.~(10)]{ArmanJ3}. Thus, for a mobile transmitter and a mobile receiver, the impulse response of the channel, denoted by $h(t,\tau)$, can be written as 
\begin{IEEEeqnarray}{C}
	\label{Eq.CIRTimVar}
	h(t,\tau) = \frac{V_{\mathrm{obs}}}{(4\pi D_1 \tau)^{3/2}} \exp \left( \frac{- r(t)^2}{4 D_1 \tau} \right).
\end{IEEEeqnarray} 

The impulse response $h(t,\tau)$ completely characterizes the time-variant channel and is a function of both $t$ and $\tau$. Variable $t$ represents the time of release of the molecules at the transmitter, whereas $\tau$ represents the relative time of observation of the signaling molecules at the receiver for a fixed value of $t$, cf. Fig.~\ref{Fig.CIR}. We note that the movement of the receiver is accounted for in \eqref{Eq.CIRConditioned} via $D_1$ as far as its effect on the $A$ molecules is concerned, and in \eqref{Eq.PDFr(t)} via $D_2$ as far as the relative motion of the transmitter and receiver is concerned. Both effects impact $h(t,\tau)$ in \eqref{Eq.CIRTimVar}. For any given $\tau$, $h(t,\tau)$ is a stochastic process with random variables $h(t_i, \tau), i \in \{1,2, \ldots, n\}$. Specifically, $h(t_i, \tau)$ can be interpreted as a function of random variable $\vec{r}(t)$.      

\subsection{Statistical Averages of Time-Variant MC Channel}
In this subsection, we analyze the statistical averages of the considered time-variant channel, i.e., the statistical averages of the random process $h(t,\tau)$. In particular, we derive closed-form analytical expressions for the mean and autocorrelation function of $h(t,\tau)$. In the remainder of this paper, for conciseness of presentation, we introduce the following notations: 
\begin{IEEEeqnarray}{C}
	\label{Eq.Notations}
	\varphi \hs = \frac{V_{\mathrm{obs}}}{(4 \pi D_1 \tau)^{3/2}},\, \lambda(t) \hs = \frac{1}{(4 \pi D_2 t)^{3/2}},\,
	\alpha \hs = \frac{1}{4 D_1 \tau},\, \beta(t) \hs = \frac{1}{4 D_2 t}. \nonumber \\*
\end{IEEEeqnarray}

Let us start with the mean of $h(t,\tau)$ for arbitrary time $t$, $m(t)$. Then, $m(t)$  can be evaluated as  
\begin{IEEEeqnarray}{C}
	\label{Eq.mean_h_def}
	m(t) = \mathcal{E}\left\lbrace h(t,\tau)\right\rbrace = \hs \hs \int \limits_{\vec{r} \in \mathbb{R}^3} h(t,\tau)\bigg|_{\vec{r}(t) = \vec{r}} \hs \hs \times \textit{{\large f}}_{\vec{r}(t)}(\vec{r}) \dif \vec{r},
\end{IEEEeqnarray}
where $\mathcal{E}(\cdot)$ denotes expectation. The solution to \eqref{Eq.mean_h_def} is provided in the following theorem.
\begin{theorem}[Mean of Time-variant MC Channel] 
The mean of a time-variant MC channel consisting of diffusive passive transmitter and receiver nano-machines with diffusion coefficients $\diffd{\TX}$ and $\diffd{\RX}$, respectively, which communicate via signaling molecules with diffusion coefficient $\diffd{A}$, is given by
\begin{IEEEeqnarray}{C}
	\label{Eq.mean_h}
\hspace{-4 mm}	m(t) = \frac{V_{\mathrm{obs}}}{\left( 4\pi \left(D_1 \tau + D_2 t \right) \right)^{3/2}} \exp \left( \frac{-x_0^2}{4 \left(D_1 \tau + D_2 t \right) } \right).
\end{IEEEeqnarray}
\end{theorem}
\begin{IEEEproof} 
Substituting \eqref{Eq.PDFr(t)} and \eqref{Eq.CIRTimVar} in \eqref{Eq.mean_h_def}, we can write $m(t)$ as 
\begin{IEEEeqnarray}{rCl} 
	\label{Eq.mean_h_expand}
	m(t) & = & \varphi \lambda(t) \int \limits_{\vec{r} \in \mathbb{R}^3} e^{-\alpha |\vec{r}|^2 } \times  e^{ - \beta(t) |\vec{r} - \vec{r}_0|^2 } \dif \vec{r}. \nonumber \\
	& = & \varphi \lambda(t) \int \limits_{-\infty}^{+\infty} \int \limits_{-\infty}^{+\infty} \int \limits_{-\infty}^{+\infty} e^{ -\left(\alpha + \beta(t)\right)x^2 + 2\beta(t)x_0x - \beta(t)x_0^2 } \nonumber \\
	&& \times\> e^{ -\left(\alpha + \beta(t)\right)y^2 } \times e^{ -\left(\alpha + \beta(t)\right)z^2 } \dif x \dif y \dif z.
\end{IEEEeqnarray}
Now, \hs using the following definite integral \cite[Eq.~(3.323.2.10)]{Gradshteyn} 
\begin{IEEEeqnarray}{C} 
	\label{Eq.Integral} 
 \int \limits_{-\infty}^{+\infty} \exp \left( -p^2 x^2  \pm qx \right) \dif x = \exp \left(  \frac{q^2}{4p^2}\right) \frac{\sqrt{\pi}}{p},
\end{IEEEeqnarray}
the integrals in \eqref{Eq.mean_h_expand} simplify to the expression in \eqref{Eq.mean_h}.
\end{IEEEproof} 

\begin{remark}
Since $m(t)$ is a function of $t$, $h(t,\tau)$ is a non-stationary stochastic process. In fact, this is due to the assumption of an unbounded environment, as on average the transmitter and the receiver diffuse away from each other and, ultimately, $h(t,\tau)$ approaches zero as $t \to \infty$.
\end{remark} 

Next, we derive a closed-form expression for the \emph{autocorrelation function} (ACF) of $h(t,\tau)$ for two arbitrary times $t_1$ and $t_2 > t_1$, denoted as $\phi(t_1,t_2)$. To this end, we write $\phi(t_1,t_2)$ as follows \footnote{In our analysis, the definition of the ACF in \eqref{Eq.ACF_def} can be easily extended to $\phi(t_1,t_2) = \mathcal{E}\left\lbrace h(t_1,\tau_1) h(t_2,\tau_2) \right\rbrace$. However, since we consider a detector that takes only one sample at a fixed time after the beginning of each modulation interval, we focus on the simplified case of $\tau_1 = \tau_2 = \tau$.}
\begin{IEEEeqnarray}{rCl}
	\label{Eq.ACF_def}
	\phi(t_1,t_2) & = & \mathcal{E}\left\lbrace h(t_1,\tau) h(t_2,\tau)\right\rbrace = \iint \limits_{\vec{r}_1,\,\vec{r}_2 \in \mathbb{R}^3} h(t_1,\tau)\big|_{\vec{r}(t) = \vec{r}_1} \nonumber \\
	&& \times\> h(t_2,\tau)\big|_{\vec{r}(t) = \vec{r}_2} \times \textit{{\large f}}_{\vec{r}(t_1),\,\vec{r}(t_2)} \left(\vec{r}_1,\, \vec{r}_2 \right) \dif \vec{r}_1 \dif \vec{r}_2, 
\end{IEEEeqnarray} 
where $\textit{{\large f}}_{\vec{r}(t_1),\,\vec{r}(t_2)}\left(\vec{r}_1,\, \vec{r}_2 \right)$ is the joint distribution function of random variables $\vec{r}(t_1)$ and $\vec{r}(t_2)$, which can be written as 
\begin{IEEEeqnarray}{C} 
	\label{Eq.jointPDF}
	\textit{{\large f}}_{\vec{r}(t_1),\,\vec{r}(t_2)}\left(\vec{r}_1,\, \vec{r}_2 \right) = \textit{{\large f}}_{\vec{r}(t_1)}\left( \vec{r}_1 \right) \textit{{\large f}}_{\vec{r}(t_2)} \left( \vec{r}_2 \, \big| \, \vec{r}_1 \right),   
\end{IEEEeqnarray}
where we used the fact that free diffusion is a memoryless process and, as a result, $\textit{{\large f}}_{\vec{r}(t_2)} \left( \vec{r}_2 \, \big| \, \vec{r}_1,\, \vec{r}_0 \right) = \textit{{\large f}}_{\vec{r}(t_2)} \left( \vec{r}_2 \, \big| \, \vec{r}_1 \right)$. Given \eqref{Eq.jointPDF}, the solution to $\phi(t_1,t_2)$ is provided in the following theorem.

\begin{theorem}[ACF of Time-variant MC Channel] 
The ACF of the impulse response of a time-variant MC channel consisting of diffusive passive transmitter and receiver nano-machines with diffusion coefficients $\diffd{\TX}$ and $\diffd{\RX}$, respectively, which communicate via signaling molecules with diffusion coefficient $\diffd{A}$, is given by
\begin{IEEEeqnarray}{C}
	\label{Eq.ACF}
\hspace{-5 mm}	\phi(t_1,t_2) = \frac{(2\pi)^3 \varphi^2 \lambda(t_1) \lambda(t_2 - t_1) \exp \left( \kappa \right)}{\left(4\left(\alpha \hs + \hs \beta \left(t_1\right) \right) \hs \left( \alpha \hs + \hs \beta \left( t_2 \hs - \hs t_1 \right) \right) \hs + \hs \alpha \beta \left( t_2 \hs - \hs t_1 \right)\right)^{3/2}},
\end{IEEEeqnarray} 
where $t_1$ and $t_2 > t_1$ are two arbitrary times and $\kappa$ is defined as 
\begin{IEEEeqnarray}{C}
	\label{Eq.ACF_kappa} 
	\kappa = \frac{- \alpha \beta \left( t_1 \right)x_0^2}{\left( \alpha + \beta \left( t_1 \right) \right) \left( \alpha + \beta \left( t_2 - t_1 \right) \right) + \alpha \beta \left( t_2 - t_1 \right)}.
\end{IEEEeqnarray}  
\end{theorem}
\begin{IEEEproof}
Please refer to the Appendix.
\end{IEEEproof} 

In the following corollary, we study a special case of $\phi(t_1,t_2)$ where $t_2 \to t_1$, i.e., $\phi(t_1, t_1) = \mathcal{E}\left\lbrace h(t_1,\tau) h(t_1,\tau)\right\rbrace$, since $\phi(t_1,t_1)$ cannot be directly obtained from \eqref{Eq.ACF} after substituting $t_2 = t_1$. 
\begin{corollary}[ACF of Time-variant MC Channel for $t_2 = t_1$] In the limit of $t_2 \to t_1$, the ACF of $h(t, \tau)$, i.e., $\phi(t_1, t_1)$, is given by  
\end{corollary}
\begin{IEEEeqnarray}{C}
	\label{Eq.ACF_equaltime}
	\phi(t_1,t_1) = \frac{V_{\mathrm{obs}}^2 \exp \left( \frac{-x_0^2}{ 2 \left(D_1 \tau + 2 D_2 t_1 \right)} \right)}{\left( 4 \pi D_1 \tau \right)^{3/2} \left( 4\pi \left(D_1 \tau + 2 D_2 t_1 \right) \right)^{3/2}}. 
\end{IEEEeqnarray}
\begin{IEEEproof}
In the limit of $t_2 \to t_1$, $\phi(t_1,t_2)$ in \eqref{Eq.ACF_def} becomes
\begin{IEEEeqnarray}{C} 
	\label{ACF_equaltime_proof_def}
	\phi(t_1,t_1) = \mathcal{E}\left\lbrace h^2(t_1,\tau) \right\rbrace = \int \limits_{\vec{r}_1 \in \mathbb{R}^3} h^2(t_1,\tau)\bigg|_{\vec{r}(t) = \vec{r}_1} \hs \hs \times \textit{{\large f}}_{\vec{r}(t_1)}(\vec{r}_1) \dif \vec{r}_1. \nonumber \\*
\end{IEEEeqnarray}
Substituting \eqref{Eq.PDFr(t)} and \eqref{Eq.CIRTimVar} in \eqref{ACF_equaltime_proof_def}, leads to 
\begin{IEEEeqnarray}{C} 
	\label{Eq.ACF_equaltime_proof_sub} 
 \hspace{-4 mm}	\phi(t_1,t_1) = \varphi^2 \lambda(t_1) \int \limits_{\vec{r}_1 \in \mathbb{R}^3} e^{ -2 \alpha |\vec{r}_1|^2 } \times  e^{ - \beta(t_1) |\vec{r}_1 - \vec{r}_0|^2 } \dif \vec{r}_1.
\end{IEEEeqnarray} 
Now, expanding the integrand in \eqref{Eq.ACF_equaltime_proof_sub}, similar to \eqref{Eq.mean_h_expand}, and using \eqref{Eq.Integral}, $\phi(t_1,t_1)$ simplifies to \eqref{Eq.ACF_equaltime}.  
\end{IEEEproof} 
Given \eqref{Eq.ACF_equaltime}, we define the variance of the time-variant MC channel as $\sigma^2(t) = \phi(t,t) - m^2(t)$.  
\subsection{Coherence Time of Time-Variant MC Channel} 
In this subsection, we provide an expression for evaluation of the coherence time of the considered time-variant MC channel. To this end, we first define the normalized autocorrelation function of random process $h(t,\tau)$ as follow: 
\begin{IEEEeqnarray}{C}
	\label{Eq.CorrCoef_def}
	\rho(t_1,t_2) \hs = \frac{\mathcal{E}\left\lbrace h(t_1,\tau) h(t_2,\tau)\right\rbrace}{\sqrt{\mathcal{E}\left\lbrace h^2(t_1,\tau)\right\rbrace \mathcal{E}\left\lbrace h^2(t_2,\tau)\right\rbrace}} = \frac{\phi(t_1,t_2)}{\sqrt{\phi(t_1,t_1)\phi(t_2,t_2)}}. \nonumber \\*
\end{IEEEeqnarray}
Now, for time $t_1 = 0$, we define the coherence time of the time-variant MC channel, $T^{\text{c}}$, as the minimum time $t_2$ after $t_1 = 0$ for which $\rho(t_1,t_2)$ falls below a certain threshold value $0 < \eta < 1 $, i.e., 
\begin{IEEEeqnarray}{C}
	\label{Eq.CoherenceTime_def}
	T^{\text{c}} = \inf \limits_{\forall t_2 > 0} \left( \rho(0,t_2) < \eta \right).
\end{IEEEeqnarray}
We note that the particular choice of $\eta$ is application dependent and may vary from one application scenario to another. For example, typical values of $\eta$ reported in the traditional wireless communication systems literature cover the full range of $0$ to $1$, see e.g. \cite{Giannakis,Vicario,Wang}. 
\section{Error Rate Analysis for Perfect and Outdated CSI}
\label{Sec.PerAna}
In this section, we first calculate the expected error probability of a single-sample threshold detector. Then, we provide a discussion on the choice of the detection threshold of the detector. Finally, in order to investigate the impact of CIR decorrelation, we look at the expected error probability of the considered detector for perfect and outdated CSI.

\subsection{Expected Bit Error Probability}
We consider a single-sample threshold detector, where the receiver takes one sample at a fixed time $\tau_s$ after the release of the molecules at the transmitter in each modulation bit interval, counts the number of signaling $A$ molecules inside its volume, and compares it with a detection threshold. In particular, we denote the received signal, i.e., the number of observed molecules inside the volume of the receiver, in the $j$th bit interval, $j \in \{1,2,\ldots, L\}$, at the time of sampling by $N(\tau_{j,s})$, where $\tau_{j,s} = (j-1)T + \tau_s$. Furthermore, we assume that the detection threshold of the receiver can be adapted from one bit interval to the next, and we denote it by $\xi_j$. The choice of $\xi_j$ is discussed in the next subsection. Thus, the decision of the single-sample detector in the $j$th bit interval, $\hat{\bit}_j$, is given by
\begin{equation}
	\label{Eq.Reception} 
	\hat{\bit}_j = \begin{cases} 
	1 &\mbox{if } N(\tau_{j,s}) \geq \xi_j, \\
	0 &\mbox{if } N(\tau_{j,s}) < \xi_j. 
			\end{cases}
\end{equation}

For the decision rule in \eqref{Eq.Reception}, we showed in \cite{ArmanJ3} that the expected error probability of the $j$th bit, $\mathrm{\overline{P}_e}(b_j)$, can be calculated as \cite[Eq.~(12)]{ArmanJ3}
\begin{IEEEeqnarray}{C}
	\label{Eq.BER_bj}
 \hspace{-6mm}	\mathrm{\overline{P}_e}(b_j) = \idotsint\limits_{\mathbf{r} \in \mathcal{R}} \sum_{\seq \in \mathcal{B}} \textit{{\large f}}_{\vec{R}}\left( \mathbf{r} \right) \pr(\seq) \Pe(\bit_j | \seq, \mathbf{r}) \dif \vec{r}_1 \cdots \dif \vec{r}_{L-1}, 
\end{IEEEeqnarray}
where $\textit{{\large f}}_{\vec{R}}\left( \mathbf{r} \right)$ is the ($L-1$)-dimensional joint PDF of vector $\vec{R} = [\vec{r}(T), \vec{r}(2T), \cdots, \vec{r}((L-1)T)]$ that can be evaluated as      
\begin{IEEEeqnarray}{rCl}
	\label{Eq.JointPDF}
	\textit{{\large f}}_{\vec{R}}\left( \mathbf{r} \right) & = & \textit{{\large f}}_{\vec{r}(T)}\left( \vec{r}_1 | \vec{r}_{0} \right) \times \cdots \times \textit{{\large f}}_{\vec{r}\left((L-1)T\right)}\left( \vec{r}_{L-1} | \vec{r}_{L-2}, \cdots, \vec{r}_0 \right) \nonumber \\
	 & \overset{(a)}{=} & \prod \limits_{j = 1}^{j=L-1} \textit{{\large f}}_{\vec{r}(jT)}\left( \vec{r}_j | \vec{r}_{j-1} \right).
\end{IEEEeqnarray}
Here, $\mathbf{r} = [\vec{r}_1, \vec{r}_2, \cdots, \vec{r}_{L-1}]$ is one sample realization of $\vec{R}$ and equality $(a)$ holds as free diffusion is a memoryless process, i.e., $\textit{{\large f}}_{\vec{r}(jT)}\left( \vec{r}_j | \vec{r}_{j-1}, \cdots, \vec{r}_0 \right) = \textit{{\large f}}_{\vec{r}(jT)}\left( \vec{r}_j | \vec{r}_{j-1} \right)$. Furthermore, $\mathcal{R}$ and $\mathcal{B}$ are the sets containing all possible realizations of $\mathbf{r}$ and $\seq$, respectively, and $\pr(\seq)$ denotes the likelihood of the occurrence of $\seq$ and $\Pe(\bit_j | \seq, \mathbf{r})$ is the conditional bit error probability of $\bit_j$. In \cite{ArmanJ3}, we considered a \emph{reactive receiver} and showed how $\Pe(\bit_j | \seq, \mathbf{r})$ can be calculated for a single-sample detector using a \emph{fixed} detection threshold $\xi$. Here, we provide $\Pe(\bit_j | \seq, \mathbf{r})$ for a \emph{transparent receiver} employing a single-sample detector with an \emph{adaptive} detection threshold $\xi_j$. 

Let us assume that $\seq$ and $\mathbf{r}$ are known. It has been shown in \cite{NoelJ1} that the number of observed molecules, $N(\tau_{j,s})$, can be accurately approximated by a Poisson random variable. The mean of $N(\tau_{j,s})$, denoted by $\overline{N}(\tau_{j,s})$, due to the transmission of all bits up to the current bit interval can be written as 
\begin{IEEEeqnarray}{C} 
	\label{Eq.MeanReceivedSig} 
\hspace{-4mm}	\overline{N}(\tau_{j,s}) = N_{A} \sum_{i=1}^{j} \bit_i h \left( iT, (j-i)T +\tau_{s} \right)\bigg|_{\vec{r}(iT) = \vec{r}_i} + \overline{n}_A,
\end{IEEEeqnarray} 
where $\overline{n}_A$ is the mean number of noise molecules inside the volume of the receiver at any given time. Now, given $\overline{N}(\tau_{j,s})$ and the decision rule in \eqref{Eq.Reception}, $\Pe(\bit_j | \seq, \mathbf{r})$ can be written as 
\begin{IEEEeqnarray}{C}
	\label{Eq.BER_bj_Conditional}
	\Pe(\bit_j | \seq, \mathbf{r}) = \begin{cases} 
	\pr(N(\tau_{j,s}) < \xi_j) & \mbox{if } \bit_j = 1, \\
	\pr(N(\tau_{j,s}) \geq \xi_j) &\mbox{if } \bit_j = 0, 
			\end{cases} 
\end{IEEEeqnarray}
where $\pr(N(\tau_{j,s}) < \xi_j)$ can be calculated from the cumulative distribution function of a Poisson distribution as 
\begin{IEEEeqnarray}{C} 
	\label{Eq.Poisson_CDF}
	\pr(N(\tau_{j,s}) < \xi_j)  =  \exp\left(-\overline{N}(\tau_{j,s})\right) \sum_{\omega = 0}^{\xi_j - 1} \frac{\left(\overline{N}(\tau_{j,s})\right)^{\omega}}{\omega !},
\end{IEEEeqnarray}
and $\pr(N(\tau_{j,s}) \geq \xi_j) = 1 - \pr(N(\tau_{j,s}) < \xi_j)$. Given $\mathrm{P_e}(\bit_j| \seq, \mathbf{r})$ in \eqref{Eq.BER_bj_Conditional}, $\mathrm{\overline{P}_e}(\bit_j)$ can be calculated based on \eqref{Eq.BER_bj}. Subsequently, we can write $\mathrm{\overline{P}_e} = \frac{1}{L} \sum_{j=1}^{L} \mathrm{\overline{P}_e}(\bit_j)$.   
\subsection{Choice of Detection Threshold} 
In this subsection, we provide a discussion regarding the choice of the adaptive detection threshold for the considered single-sample detector. Let us assume for the moment that sequence $\{\bit_1, \bit_2, \cdots, \bit_{j-1}\}$ and $\mathbf{r}$ are known, and we are interested in finding the optimal detection threshold, $\xi_j^{\text{opt}}$, that minimizes instantaneous error probability $\mathrm{P_e}(\bit_j)$. Then, we have shown in \cite{ArmanJ1} that for any threshold detector whose received signal can be modeled as a Poisson random variable, $\xi_j^{\text{opt}}$ is given by \cite[Eq.~(25)]{ArmanJ1}
\begin{IEEEeqnarray}{C} 
	\label{Eq.OptimalThreshold}
	\xi_j^{\text{opt}} = \Bigg \lceil \frac{\ln \left( \frac{P_0}{P_1} \right) + \left( \lambda_1 - \lambda_0 \right)}{\ln \left( \lambda_1 / \lambda_0 \right)} \Bigg \rceil,
\end{IEEEeqnarray} 
where $\lambda_1 = \overline{N}(\tau_{j,s}|\bit_j = 1)$,  $\lambda_0 = \overline{N}(\tau_{j,s}|\bit_j = 0)$, and $\lceil \cdot \rceil$ denotes the ceiling function. 
\begin{remark}
We note that the evaluation of $\xi_j^{\text{opt}}$ requires knowledge of the previously transmitted bits up to the current bit interval, which is not available in practice. Thus, for practical implementation, we propose a suboptimal detector whose detection threshold, $\xi_j^{\text{Subopt}}$, is evaluated according to \eqref{Eq.OptimalThreshold} after replacing $\{ \bit_1, \bit_2, \ldots \bit_{j-1}\}$ with the estimated previous bits, i.e., $\{\hat{\bit}_1, \hat{\bit}_2, \cdots, \hat{\bit}_{j-1}\}$.      
\end{remark} 
\begin{remark}
It has been shown in \cite{JamaliC1} that when the effect of inter-symbol interference (ISI) is negligible compared to $\overline{n}_A$, the combination of \eqref{Eq.Reception} and \eqref{Eq.OptimalThreshold} constitutes the optimal maximum likelihood (ML) detector. We note that, in this regime, knowledge of previously transmitted bits is not required for calculation of $\xi_j^{\text{opt}}$.
\end{remark} 
\subsection{Detectors with Perfect and Outdated CSI} 
In this subsection, we distinguish between two cases regarding the CSI knowledge, namely \emph{perfect} CSI and \emph{outdated} CSI, and explain how the corresponding expected error probabilities of the single-sample detector can be evaluated. In particular, for the problem at hand, knowledge of the CSI is equivalent to knowledge of the CIR. The analytical expression for the CIR of an MC channel depends on the environment under consideration, and may not always be available in closed form. However, for our system model in Section~\ref{Sec.SysMod}, the CIR can be expressed in closed-form as in \eqref{Eq.CIRTimVar}.

\textit{Perfect CSI:} For the case of a single-sample detector with perfect CSI, we assume that for any given modulation bit interval, $r(t)$ is known at the receiver for all previous bit intervals up to the current bit interval, i.e., at the $j$th bit interval, $[\vec{r}(0), \vec{r}(T), \ldots, \vec{r}(jT)]$ is known at the receiver. Thus, $\xi_j^{\text{Subopt}}$ can be directly obtained from \eqref{Eq.CIRTimVar}, \eqref{Eq.MeanReceivedSig}, and \eqref{Eq.OptimalThreshold}.

\textit{Outdated CSI:} For the case of a single-sample detector with outdated CSI, we assume that \emph{only} the initial distance between transmitter and receiver at time $t_0 = 0$, i.e., $r_0$, is known at the receiver. As a result, in any modulation bit interval, the receiver evaluates $\xi_j^{\text{Subopt}}$ via \eqref{Eq.OptimalThreshold} with the mean given by \vspace{-1 mm} 
\begin{IEEEeqnarray}{C} 
	\label{Eq.MeanOutDatedCSI} 
\hspace{-4mm}	\overline{N}(\tau_{j,s}) = N_{A} \sum_{i=1}^{j} \bit_i h \left( t_0\,, (j-i)T +\tau_{s} \right) + \overline{n}_A.
\end{IEEEeqnarray}
 
Finally, for both cases, $\Pe(\bit_j | \seq, \mathbf{r})$ is obtained from
\eqref{Eq.BER_bj_Conditional}.\vspace{-1 mm}           
\section{Simulation Results}
\label{Sec.SimRes}
\begin{table}
	\renewcommand{\arraystretch}{1.20} 
	\centering 
	\caption{Simulation Parameters}\vspace{-2mm}
	\begin{tabular}{|c|c|  |c|c|}\hline 
      Parameter & Value & Parameter & Value \\ \hline \hline
        $N_A$ & $30000$ & $T$ & $0.5$ ms  \\ \hline
        $D_A$ & $5 \times 10^{-9}$ $\text{m}^2/{\text{s}}$ & $\tau = \tau_s$ & $0.035$ ms \\ \hline
        $D_{\text{rx}}$ & $10^{-13}$ $\text{m}^2/{\text{s}}$ & $L$   & $50$ \\ \hline
        $r_0$ & $1$\, $\mu$m & $P_1$   & $0.5$\\ \hline
        $a_{\text{rx}}$   & $0.15$\, $\mu$m &  $P_0$ & 0.5  \\ \hline
        $\overline{n}_A$   & $10$ &  $\Delta t$ & 5 $\mu$s \\ \hline
      \end{tabular} 
      \label{Table.1}
\end{table}
In this section, we present simulation and analytical results to assess the accuracy of the derived analytical expressions for the mean and the ACF of the time-variant CIR and the expected error probability of the considered mobile MC system. 
For simulation, we adopted a particle-based simulation of Brownian motion, where the precise locations of the signaling molecules, transmitter, and receiver are tracked throughout the simulation environment. In particular, in the simulation algorithm, time is advanced in discrete steps of $\Delta t$ seconds. In each step of the simulation, each $A$ molecule, the transmitter, and the receiver undergo random walks, and their new positions in each Cartesian coordinate are obtained by sampling a Gaussian random variable with mean $0$ and standard deviation $\sqrt{2D_A\Delta t}$, $\sqrt{2 \diffd{\TX} \Delta t}$, and $\sqrt{2 \diffd{\RX} \Delta t}$, respectively. Furthermore, we used Monte-Carlo simulation for evaluation of the multi-dimensional integral in \eqref{Eq.BER_bj}.   

For all simulation results, we chose the set of simulation parameters provided in Table~\ref{Table.1}, unless stated otherwise. Furthermore, we considered an environment with the viscosity of blood plasma ($\simeq 4\,\text{mPa}\cdot\text{s}$) at $37\, \mathrm{{}^{\circ}C}$ and we used the Stokes--Einstein equation \cite[Eq.~(5.7)]{NakanoB} for calculation of $D_A$ and $\diffd{\TX}$. The only parameter that was varied is $\diffd{\TX}=\{0.1, 1, 5, 20, 100\}\times 10^{-13}$ $\frac{\text{m}^2}{\text{s}}$ (corresponding to $a_{\text{tx}} = 5.6793 \times \{10^{-6}, 10^{-7}, 2 \times 10^{-8}, 5 \times 10^{-9}, 1 \times 10^{-9} \}$ m).\footnote{Small values of $a_{\text{tx}}$ (in the order of a few nm) have been used only to consider the full range of $\diffd{\TX}$ values.} All simulation results were averaged over $10^{5}$ independent realizations of the environment.  

\begin{figure}[!t]
	\centering
	\includegraphics[scale=0.35]{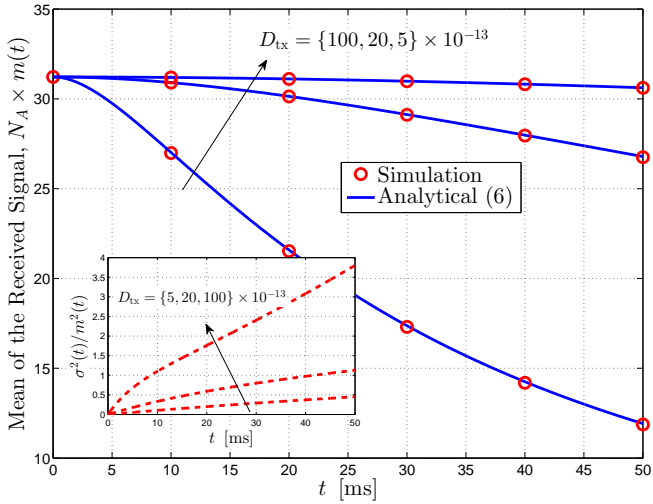}
	\caption{Expected received signal in the absence of external noise molecules, $N_A m(t)$, as a function of time $t$.}
	\label{Fig.Analysis2}
\end{figure}

\begin{figure}[!t]
	\centering
	\includegraphics[scale=0.35]{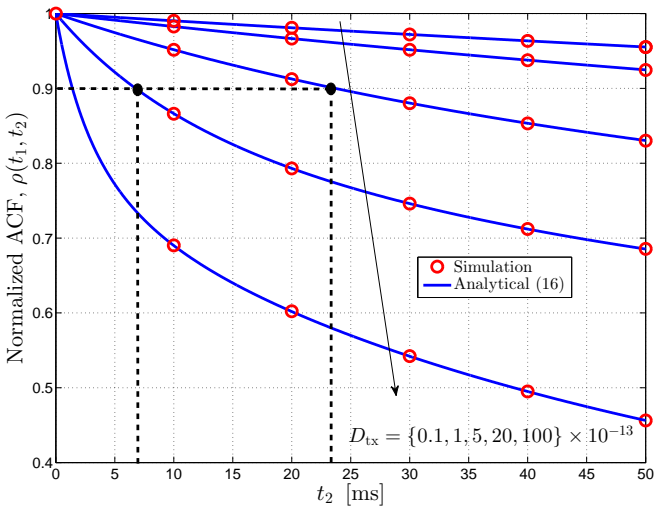}
	\caption{Normalized ACF, $\rho(t_1,t_2)$, as a function of $t_2$, for a fixed $t_1 = 0$, $\tau = \tau_s$, and different values of $\diffd{\TX}=\{0.1, 1, 5, 20, 100\}\times 10^{-13}$ $\frac{\text{m}^2}{\text{s}}$.}
	\label{Fig.Analysis1}
\end{figure}
In Fig.~\ref{Fig.Analysis2} and its inset, we investigate the impact of increasing time $t$ on the mean and the normalized variance of the received signal in the absence of external noise molecules, i.e., $N_Am(t)$ and $\sigma^2(t) / m^2(t)$, respectively, for system parameters $\diffd{\TX}=\{5, 20, 100\}\times 10^{-13}$ $\frac{\text{m}^2}{\text{s}}$. Fig.~\ref{Fig.Analysis2} shows that as time $t$ increases, $N_A m(t)$ decreases. This is due to the fact that as $t$ increases, on average $r(t)$ increases as transmitter and receiver diffuse away and, consequently, $m(t)$ decreases. The decrease is faster for larger values of $\diffd{\TX}$, since for larger $\diffd{\TX}$, the transmitter diffuses away faster. The normalized variance of the received signal is shown in the inset of Fig.~\ref{Fig.Analysis2}. We observe that for all values of $\diffd{\TX}$, the normalized variance of the received signal is an increasing function of time. This is because as time increases, due to the Brownian motion of the transmitter and the receiver, the variance of the movements of both of them increases, which leads to an increase in the normalized variance of the received signal. As expected, this increase is faster for larger values of $\diffd{\TX}$, since the displacement variance of the transmitter, $2\diffd{\TX}t$, is larger.

\begin{figure}[!t]
	\centering
	\includegraphics[scale=0.35]{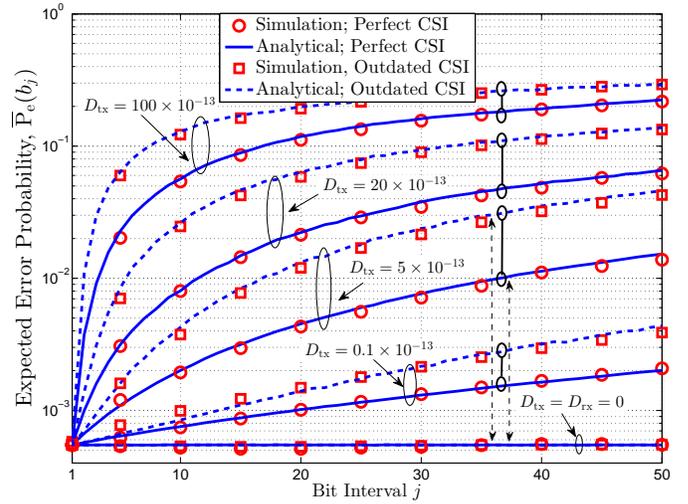}
	\caption{Expected error probability, $\mathrm{\overline{P}_e}(b_j)$, as a function bit interval $j$.}
	\label{Fig.Analysis3}
\end{figure}
\begin{figure*}[!t]
\small 
\setcounter{tempeqcounter}{\value{equation}} 
\begin{IEEEeqnarray}{rCl}
\setcounter{equation}{27} 
	\label{Eq.ACF_Proof2} 
\phi(t_1,t_2) & = & \varphi^2 \lambda(t_2 - t_1) \lambda(t_1) \int_{-\infty}^{+\infty} \cdots \int_{-\infty}^{+\infty} \exp\left(  -\left( \alpha + \beta(t_2 - t_1) + \beta(t_1) \right)z_1^2 - \left( \alpha + \beta(t_2 - t_1) \right)z_2^2 + 2\beta(t_2 - t_1)z_1z_2  \right) \nonumber \\
&&  \times\>  \exp\left(  -\left( \alpha + \beta(t_2 - t_1) + \beta(t_1) \right)y_1^2 - \left( \alpha + \beta(t_2 - t_1) \right)y_2^2 + 2\beta(t_2 - t_1)y_1y_2  \right) \times \exp \left( -\left( \alpha + \beta(t_2 - t_1) + \beta(t_1) \right)x_1^2 \right. \nonumber \\ 
&&  -\> \left. \left( \alpha + \beta(t_2 - t_1) \right)x_2^2 + 2\beta(t_2 - t_1)x_1x_2 + 2\beta(t_1)x_0x_1 - \beta(t_1)x_0^2 \right) \dif z_1 \dif z_2 \dif z_3 \dif y_1 \dif y_2 \dif y_3 \dif x_1 \dif x_2 \dif x_3.   	 
\end{IEEEeqnarray}
\setcounter{equation}{\value{tempeqcounter}} 
\hrulefill
\vspace*{4pt} 
\vspace*{-6 mm}
\end{figure*}  

In Fig.~\ref{Fig.Analysis1}, the normalized ACF, $\rho(t_1,t_2)$, is evaluated as a function of $t_2$ for a fixed value of $t_1 = 0$ and transmitter diffusion coefficients $\diffd{\TX}=\{0.1, 1, 5, 20, 100\}\times 10^{-13}$ $\frac{\text{m}^2}{\text{s}}$. We observe that for all considered values of $\diffd{\TX}$, $\rho(t_1,t_2)$ decreases with increasing $t_2$. This is due to the fact that for increasing $t_2$, on average $r(t)$ increases, and the CIR becomes more decorrelated from the CIR at time $t_1 = 0$. Furthermore, as expected, for larger values of $\diffd{\TX}$, $\rho(t_1,t_2)$ decreases faster, as for larger values of $\diffd{\TX}$, the transmitter diffuses away faster. We can see that, e.g., for the choice of $\eta = 0.9$, the coherence time, $T^{\text{c}}$, for $\diffd{\TX} = 20 \times 10^{-13}$ $\frac{\text{m}^2}{\text{s}}$ and $\diffd{\TX} = 5 \times 10^{-13}$ $\frac{\text{m}^2}{\text{s}}$ is $7$ ms and $23$ ms, respectively.
 
In Fig.~\ref{Fig.Analysis3}, the expected error probability, $\mathrm{\overline{P}_e}(b_j)$, is shown as a function of bit interval $j$, for system parameters $\diffd{\TX}=\{0.1, 5, 20, 100\}\times 10^{-13}$ $\frac{\text{m}^2}{\text{s}}$, as well as for fixed transmitter and receiver, i.e., $\diffd{\TX} = \diffd{\RX} = 0$. As expected, when transmitter and receiver are fixed, the performances of the detectors with perfect and outdated CSI are identical, as the channel does not change over time. On the other hand, when $\diffd{\TX} > 0$ and/or $\diffd{\RX} > 0$, the performance of both detectors deteriorates over time. This is due to the fact that as time increases, i) $\sigma^2(t)$ increases and ii) $m(t)$ decreases. However, we can also observe that the gap between the BERs of the detector with perfect CSI and the detector with outdated CSI increases over time since over time the impulse response of the channel decorrelates (see Fig.~\ref{Fig.Analysis1}), and, as a result, the CSI becomes outdated. Furthermore, the CSI becomes outdated faster for larger values of $\diffd{\TX}$. Hence, for a given time (bit interval), the absolute value of the performance gap between both cases, shown with black solid lines in Fig.~\ref{Fig.Analysis3}, increases. For instance, for $j = 37$, the absolute values of the performance gap between the detectors with perfect and outdated CSI for $\diffd{\TX} = \{ 0.1, 5, 20, 100\}\times 10^{-13}$ $\frac{\text{m}^2}{\text{s}}$ are $\{0.0013, 0.0212, 0.0624, 0.8\}$, respectively.

Finally, we note the excellent match between simulation and analytical results in Figs.~3-5. \vspace{-1 mm} 

\section{Conclusions and Future Work}
\label{Sec.Con} 
In this paper, we established a statistical mathematical framework for the characterization of the time-variant CIR of mobile MC channels. In particular, we derived closed-form analytical expressions for the mean and ACF of the time-variant CIR. Given the ACF, we defined the coherence time of the channel and investigated the impact that CIR decorrelation over time has on the BER performance of a single-sample detector with outdated CSI. The analysis and results in this paper reveal the necessity to design new modulation, detection, and estimation techniques for time-variant mobile MC channels. 

In this paper, we considered a simple transparent receiver. The extension of the developed mathematical framework to more sophisticated reactive and absorbing receivers is an interesting topic for future work. \vspace{-2 mm}   
\appendix[Proof of Theorem 2]
Given \eqref{Eq.jointPDF}, substituting $h(t_1,\tau)\big|_{\vec{r}(t) = \vec{r}_1}$ and $h(t_2,\tau)\big|_{\vec{r}(t) = \vec{r}_2}$ from \eqref{Eq.CIRTimVar} to \eqref{Eq.ACF_def}, we can write $\phi(t_1,t_2)$ as 
\begin{IEEEeqnarray}{rCl}
	\label{Eq.ACF_Proof1}
	\phi(t_1,t_2) & = & \varphi^2 \lambda(t_2 - t_1) \lambda(t_1) \hs \iint \limits_{\vec{r}_1,\,\vec{r}_2 \in \mathbb{R}^3} \hs e^{-\alpha |\vec{r}_1|^2}  e^{-\alpha |\vec{r}_2|^2}  e^{-\beta(t_1) |\vec{r}_1 - \vec{r}_0|^2} \nonumber \\
	&& \times\> e^{-\beta(t_2 - t_1) |\vec{r}_2 - \vec{r}_1|^2} \dif \vec{r}_2 \dif \vec{r}_1. 
\end{IEEEeqnarray} 
Expanding the integrands in \eqref{Eq.ACF_Proof1} leads to \eqref{Eq.ACF_Proof2}
%
\addtocounter{equation}{1}%
\setcounter{storeeqcounter}%
{\value{equation}}%
on top of this page. For solving the multiple integrals in \eqref{Eq.ACF_Proof2}, we use the PDF integration formula for multivariate Gaussian distributions. In particular, let us assume that vector $\mathbf{X} = [x_1,y_1,z_1,x_2,y_2,z_2]^{\intercal}$ has a multivariate Gaussian distribution with mean vector $\bm{\mu} = \mathcal{E}\{ \mathbf{X} \} \in \mathbb{R}^6$ and covariance matrix $\bm{\Sigma} = \mathcal{E} \{(\mathbf{X} - \bm{\mu})(\mathbf{X} - \bm{\mu})^\intercal \}$. Then, the well-known PDF of $\mathbf{X}$ is given by 
\begin{IEEEeqnarray}{rCl}
	\label{Eq.MultivariateGaussian}
\hspace{-5 mm} \textit{{\large f}}_{\mathbf{X}}(x_1,y_1,z_1,x_2,y_2,z_2) = \frac{\exp \left( -\frac{1}{2}  (\mathbf{X} - \bm{\mu})^\intercal \bm{\Sigma}^{-1} (\mathbf{X} - \bm{\mu})\right)}{(2\pi)^3 \sqrt{\text{det}\left( \bm{\Sigma} \right)}},
\end{IEEEeqnarray}
where $\text{det}(\cdot)$ denotes the determinant. It can be easily verified that for mean vector $\bm{\mu} = [\mu_{x_1}, 0, 0, \mu_{x_2}, 0, 0]$ and inverse covariance matrix 
\begin{IEEEeqnarray}{C}
	\label{Eq.InverseCovMat}
	\Sigma^{-1} = \begin{bmatrix}
     \vartheta & 0 & 0 & \psi & 0 & 0 \\
     0 & \vartheta & 0 & 0 & \psi & 0 \\
     0 & 0 & \vartheta & 0 & 0 & \psi \\
     \psi & 0 & 0 & \varepsilon & 0 & 0 \\
     0 & \psi & 0 & 0 & \varepsilon & 0 \\
    0 & 0 & \psi & 0 & 0 & \varepsilon 
\end{bmatrix},
\end{IEEEeqnarray}
where 
\begin{IEEEeqnarray}{C}
\vartheta = 2\left(\alpha + \beta(t_2 - t_1) + \beta(t_1)\right),\,\,\, \varepsilon = 2 \left( \alpha + \beta(t_2 - t_1) \right) \nonumber \\
\psi  = -2 \beta(t_2 - t_1), \,\,\, \mu_{x_1} = \frac{2 \beta(t_1) x_0}{\vartheta - \psi^2/\varepsilon}, \,\,\, \mu_{x_2} = \frac{-\psi \mu_{X_1}}{\varepsilon},  
\end{IEEEeqnarray} 
$\exp \left( -\frac{1}{2}  (\mathbf{X} - \bm{\mu})^\intercal \bm{\Sigma}^{-1} (\mathbf{X} - \bm{\mu})\right)\times\exp(\kappa) $ (with $\kappa$ given in \eqref{Eq.ACF_kappa}) is equal to the integrands in \eqref{Eq.ACF_Proof2}. Now, given that $\int_{-\infty}^{+\infty} \cdots \int_{-\infty}^{+\infty} \textit{{\large f}}_{\mathbf{X}}(x_1,y_1,z_1,x_2,y_2,z_2) \dif x_1 \ldots \dif z_2 = 1$, $\phi(t_1, t_2)$ can be written as 
\begin{IEEEeqnarray}{C}
	\label{Eq.ACFProof3} 
	\phi(t_1,t_2) = \varphi^2 \lambda(t_2 - t_1) \lambda(t_1) \exp(\kappa) (2\pi)^3 \sqrt{\text{det}(\bm{\Sigma)}}.
\end{IEEEeqnarray}
Given $\bm{\Sigma}^{-1}$ in \eqref{Eq.InverseCovMat}, after some calculations, it can be shown that 
\begin{IEEEeqnarray}{C}
	\label{Eq.DeterminantSol}
	\text{det}\left( \bm{\Sigma} \right) = \frac{1}{\left(\vartheta \times \varepsilon - \psi^2 \right)^3}.
\end{IEEEeqnarray}
Finally, substituting \eqref{Eq.DeterminantSol} into \eqref{Eq.ACFProof3} leads to \eqref{Eq.ACF}.                 
\bibliography{IEEEabrv,Library}
\end{document}